\documentclass[aps,showkeys,floatfix,twocolumn]{revtex4}

\usepackage{graphicx}
\usepackage{dcolumn}
\usepackage{bm}
\usepackage{color}

\begin{document}


\title[Stiffness-guided motion of a droplet on a solid
substrate]{Stiffness-guided motion of a droplet on a solid substrate} 

\author{Panagiotis E. Theodorakis} \email{panos@ifpan.edu.pl.}

\affiliation{Institute of Physics, Polish Academy of Sciences, Al.\
Lotnik\'ow 32/46, 02-668 Warsaw, Poland}
\author{Sergei E. Egorov}
 \email{sae6z@eservices.virginia.edu.}
\affiliation{Department of Chemistry, University of Virginia, Charlottesville,
VA 22901, USA}
\affiliation{ 
	Institut f\"ur Physik, Johannes Gutenberg Universit\"at Mainz, 55099 Mainz, Germany
}%
\affiliation{ 
	Leibniz-Institut f\"ur Polymerforschung, Institut Theorie der Polymere, Hohe Str. 6, 01069 Dresden, Germany
}%
\author{Andrey Milchev} \email{milchev@ipc.bas.bg}
\affiliation{Bulgarian Academy of Sciences, Institute of Physical Chemistry,
1113 Sofia, Bulgaria}%

\date{\today} 
 
\begin{abstract}
 
A range of technologies require the directed motion of nanoscale droplets on
solid substrates. A way of realizing this effect is durotaxis, whereby a 
stiffness gradient of a substrate can induce directional motion without 
requiring an energy source. Here, we report on the results of extensive 
molecular dynamics investigations of droplets on a surface with varying stiffness. 
We find that durotaxis is enhanced by increasing the stiffness gradient and,  
also, by increased wettability of the substrate, in particular, when 
droplet size decreases.  We anticipate that our study will provide further
insights into the mechanisms of nanoscale directional motion. 
\end{abstract}

                              
 
\maketitle

\section{Introduction}\label{sec:intro}
A range of technologies require the control of liquids on surfaces, such as
microfabrication \cite{Srinivasarao2001} and
coating,\cite{Chaudhury1992,Wong2011} with many of these applications also
requiring a directed motion of nanodroplets onto the
surface \cite{Lagubeau2011,Prakash2008,Chaudhury1992,Darhuber2005}. To this end,
many methods have been explored, for example heterogeneous surface chemistries
for different patterns \cite{Gau1999,Sirringhaus2000,Hanakata2015}, temperature
and electric potential gradients \cite{Chaudhury1992,Darhuber2005}, and surface
topography \cite{Lafuma2003,Courbin2007,Tretyakov2014,Tretyakov2013,
Karapetsas2016}. Other techniques for moving nanoscale objects are based on
electrical current \cite{Dundas2009,Regan2004,Zhao2010,Kudernac2011},
charge \cite{Shklyaev2013,Fennimore2003,Bailey2008}, thermal energy (selective
heating) \cite{Barreiro2008,Somada2009,Chang2010}, simple stretch \cite{Huang2014}, and complicated chemical
reactions (e.g. in biological processes) \cite{vandenHeuvel2007}. In the context
of biology, directional motion of cells takes place due to various stimuli, such
as substrate chemicals, light, gravity and electrostatic
potential \cite{Ridley2003}. Recent studies have shown that cell movements are
also guided by substrate stiffness (rigidity), a phenomenon known as
durotaxis \cite{Lo2000}.

Inspired by durotaxis in biology, solid substrate durotaxis has emerged as an
attractive research field as the motion of nanoscale objects can be guided by
substrate stiffness without the requirement for an energy source with 
implications for nanoscale actuation and energy 
conversion \cite{Style2013,Chang2015,Pham2016,Lazopoulos2008,Becton2016}. Moreover, solid 
substrate durotaxis, which was investigated by using computer simulation in the 
context of a flake sliding on a graphene substrate with a stiffness gradient, 
shares similarities with the stiffness-guided directional motion in living cells 
as in both cases weak van der Waals interactions are present \cite{Chang2015}. In 
particular, the interaction between the substrate and the flake was inversely 
proportional to the stiffness \cite{Chang2015}. Lower potential energies are more 
stable than higher ones, and this explains the motion towards the higher 
stiffness adopting a thermodynamically favorable 
state \cite{Barnard2015,Becton2016}. In this case, computer simulation has been 
an extremely useful tool to interpret the modeled process, as materials were 
electrically, chemically and thermally isolated free of defects and impurities. 
Here, we study durotaxis in the context of a liquid droplet on a solid substrate 
with a stiffness gradient by means of computer simulation. In particular, we 
explore different stiffness gradients, droplet sizes, and levels of attraction 
between the droplet and the substrate. Our results highlight the importance of 
interfacial energy between the droplet and the substrate, in agreement with 
Chang {\it et al.} \cite{Chang2015}. We anticipate that our study will provide 
further insights into substrate patterning leading to new opportunities in 
nanoscale science and technology \cite{Barnard2015}.

\section{\label{sec:methods} The Model}

In this study we consider a system comprised of a liquid droplet on a substrate 
with stiffness gradient. The droplet consists of $N_p$ polymer chains 
($N_p=100,\; 600$, or $4800$) with  $N=10$ segments each. For the chosen chain 
length the vapor pressure is sufficiently low and evaporation effects are 
negligible \cite{Tretyakov2014}. Owing to the comparatively small size of the 
droplets, important quantities that characterize the droplet, such as the 
contact angle, are subject to strong fluctuations, while in addition such 
properties become also size dependent \cite{Theodorakis2015}.

\begin{figure}
\includegraphics[scale=0.9]{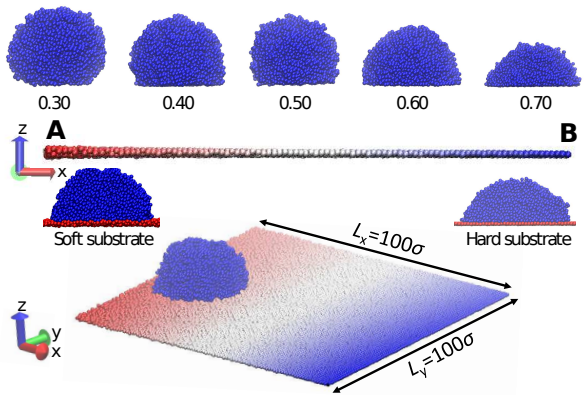}
\caption{\label{fig:1} Upper panel illustrates  the shape of the droplet
on a stiff substrate for different strength of attraction ($\varepsilon_{sp} = 
0.30 \dots  0.70$) as indicated. Middle panel shows the effect of the substrate 
stiffness gradient in $x$-direction, marked by red color (left side of 
the substrate, $A$) for soft substrate, and by blue color (on the $B$-side of 
the substrate in $x$-direction) indicating the highest rigidity. Two
examples of droplets ``sitting'' on soft and stiff substrates for the case
$\varepsilon_{sp}=0.70$ are also shown. In the bottom panel we
denote by color gradient the change in the degree of stiffness every $\Delta
L = 4\sigma$ in the $x$-direction.  Snapshots were taken by using the VMD 
software \cite{Humphrey1996}.}
\end{figure}

The substrate is  formed by a layer of spherical beads, positioned on a square
lattice with lattice constant $\alpha$, where $\alpha=\sigma$, with bead diameter
$\sigma$ serving here also as a unit of length. The substrate is chosen with
linear dimensions in the $x$ and $y$ directions $L_x = L_y = 100 \sigma$,
respectively (see Fig.~\ref{fig:1}).

Interactions between different components of the system, i.e., drop particles
and substrate beads, are described by means of the Lennard-Jones potential, 
i.e.,
\begin{equation}\label{eq:LJpotential}
U_{\rm LJ}(r) = 4\varepsilon_{\rm ij} \left[  \left(\frac{\sigma_{\rm ij}}{r}
\right)^{12} - \left(\frac{\sigma_{\rm ij}}{r}  \right)^{6}    \right],
\end{equation}
where $r$ is the distance between any pair of beads in the system, and $\rm i$
and $\rm j$ indicate the type of beads: ``$p$" stands for polymer beads, and
``$s$'' denotes substrate beads. In the present consideration, $\sigma_{\rm pp}
= \sigma_{\rm ss} = \sigma_{\rm sp} = \sigma$. As usual, the LJ-potential is cut
and shifted at a cutoff distance $r_{\rm c}=2.5\sigma$ for the ``$pp$'' and
``$sp$'' interactions, while for the interaction between substrate beads $r_{\rm
c}=2^{1/6}\sigma$ (i.e., a purely repulsive interaction). The parameter
$\varepsilon_{\rm ij}$ of the LJ potential for the polymer and the substrate is
$\varepsilon_{\rm pp} = \varepsilon_{\rm ss} = \varepsilon$, while
$\varepsilon_{\rm sp}$ was used to tune the affinity between the substrate and
the droplet. Here, $\varepsilon_{\rm sp} = 0.3$, $0.4$, $0.5$, $0.6$, and $0.7$
in units of $\varepsilon$. For example, the choice of $\varepsilon_{\rm sp} =
0.5$ provides a droplet contact-angle $\theta$ of roughly 90$^{\rm o}$ on stiff
substrates, which, of course, is size and model dependent \cite{Theodorakis2015}.

The finite extensible nonlinear elastic (FENE) potential \cite{Kremer1990} was
used to keep together consecutive beads in the polymer chain. The FENE potential
reads
 \begin{equation}\label{eq:KG}
 U_{\rm FENE}(r) = -0.5 K_{\rm FENE} R_0^2 \ln \left[ 1 -  \left(
\frac{r}{R_{\rm 0}} \right)^2    \right],
\end{equation}
where $r$ is the distance between two consecutive bonded beads along the polymer
backbone, $R_{\rm 0}=1.5\sigma$ expresses the maximum extension of the
bond, and $K_{\rm FENE} = 30 \varepsilon/\sigma^2$ is an elastic constant. 

Substrate beads are tethered to their lattice sites by a harmonic potential with
spring constant $K$ (the factor $1/2$ is absorbed in the constant), which is
used as a parameter for tuning the substrate stiffness. Small values of $K$
result in large fluctuations of the substrate beads around their position on the
lattice sites, which corresponds to small substrate stiffness (soft substrates),
whereas large values of $K$ result in strong tethering of the substrate beads to
their lattice sites resulting in a large stiffness (hard
substrate) \cite{Hanakata2015}. The stiffness gradient is implemented along the
$x$-direction (see Fig.~\ref{fig:1}): Starting from an initial stiffness
$K_{0}$, (expressed in units $\varepsilon/\sigma^2$ with $\varepsilon$ being the
unit of energy) at the very left side (point $A$ in Fig.~\ref{fig:1}) of the
substrate with coordinate $x=0$,  we increase the stiffness by $\Delta K$ every
$\Delta L = 4 \sigma$ until the right end of the substrate (point B in
Fig.~\ref{fig:1}) in the $x$-direction with stiffness $K_{\rm f}$ is reached.
Hence, the stiffness for each bead depends on the set ($K_{0}$, $\Delta K$),
according to the relation $K = K_{0} + n*\Delta K$, where $n$ is an integer that
indicates the number of times we increased the stiffness in the $x$ direction by
an amount $\Delta K$ every $\Delta L = 4 \sigma$. In our study, we explore
different sets of these parameters along with the adhesion parameter
$\varepsilon_{\rm sp}$ studying droplets of different size $N_p$.

To evolve our system in time, we used Molecular Dynamics (MD) simulations by
choosing the Langevin thermostat \cite{Schneider1978} as implemented in the
LAMMPS package \cite{Plimpton1995}. The time unit in our simulations is $\tau =
\sqrt{m\sigma^2/\varepsilon}$, where $m \equiv 1$ is the mass unit of the drop
particles and the substrate beads. The integration time-step for the
velocity-Verlet integration of the equations of motion is $\Delta t =
0.005\tau$. Thus, the temperature $T$ fluctuates around a predefined value
$T=\varepsilon/k_{B}$, where $k_{B}$ is the Boltzmann constant, and the energy
$\varepsilon$ is measured in units of $k_BT$. The total number of beads of the
system is fixed and the total volume of the system, which is also constant,
corresponds to the size of the simulation box, that is, $L_x \times L_y \times
L_z$ with $L_z$ large enough so as  to guarantee that neither the substrate nor
the polymer droplet interact with their periodic images in the $z$-direction.
Typical trajectories start at the left edge of the droplet positioned at point
$A$ of the substrate (see Fig.~\ref{fig:1}) while the simulations run up to
$10^8$MD time steps. Our results are based on the analysis of ten independent
trajectories for each set of values ($K_{0}$, $\Delta K$, $\varepsilon_{\rm
sp}$, and $N_p$). 

\section{\label{sec:results} Results}
\begin{figure}
\includegraphics[scale=0.9]{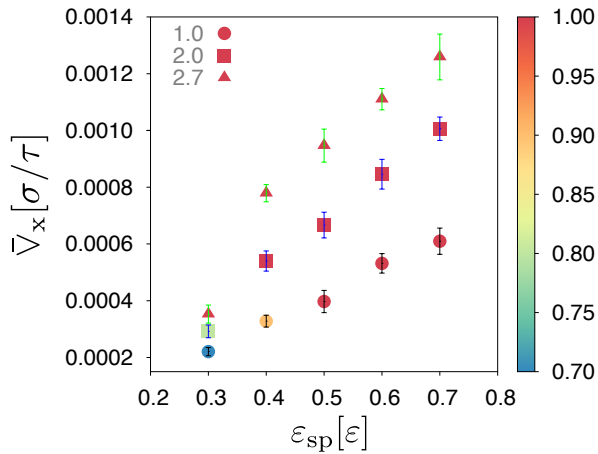}
\caption{\label{fig:2} The mean velocity $\bar{\vee}_{x}$ by which the droplet
moves from A to B (cf.Fig.~\ref{fig:1}; the center of mass of the droplet is taken
into account to estimate the distance) versus the attraction of the droplet to
the substrate $\varepsilon_{sp}$ for different values of the gradient $dK/dx$ as
indicated. The color bar indicates the probability that the droplet will cover
the distance from A to B within the simulation time ($10^8$ MD steps). The
droplet is formed by $N_p=600$ chains.}
\end{figure}

We have studied three different substrates with the {\em same} average stiffness
in order to examine the role of the stiffness gradient in durotaxis (see
Fig.~\ref{fig:2}). If the stiffness at the right most end of the substrate ($B$
in Fig.~\ref{fig:1}) is $K_{\rm f}$, then the mean stiffness will be $K_{\rm
ave} =  (K_{\rm 0} + K_{\rm f})/2$ for substrates with constant gradient along
the $x$-direction. Here, we set $K_{\rm ave} = 148 \varepsilon/\sigma^2$, and
the three chosen sets of ($K_{0}$, $\Delta K$) are $(20,10.666),\; (50, 8.166)$,
and $(100, 4)$, corresponding to values of the gradient, $K' \equiv dK/dx =
2.7,\; 2.0$, and $1.0$, respectively. The degree of droplet adhesion to the
substrate, governed by the droplet--substrate affinity $\varepsilon_{sp}$ also
affects the process of durotaxis, and should be taken into account by varying
the strength of $\varepsilon_{sp}$. With this choice of parameters, the droplet
is found to perform a predominantly translational movement from the soft edge
towards the rigid end of the substrate. In Fig.~\ref{fig:2} we plot the observed
mean velocity $\bar{\vee}_{x}$, averaged over ten individual droplets, as function of
the surface adhesion $\varepsilon_{sp}$ for the three rigidity gradients $K'$.

One should note, however, that drops, depending on the concrete values of the
aforementioned parameters, undergo partially random displacements away from the
gradient $x$-direction, so that not all droplets manage to reach the stiff edge
of the substrate within the time window of the computer experiment. Therefore,
in Fig.~\ref{fig:2} we use a color bar to indicate the share of droplets that
successfully traverse the distance between the soft and the stiff edge of the
substrate.

\begin{figure}[htb]
\includegraphics[scale=0.95]{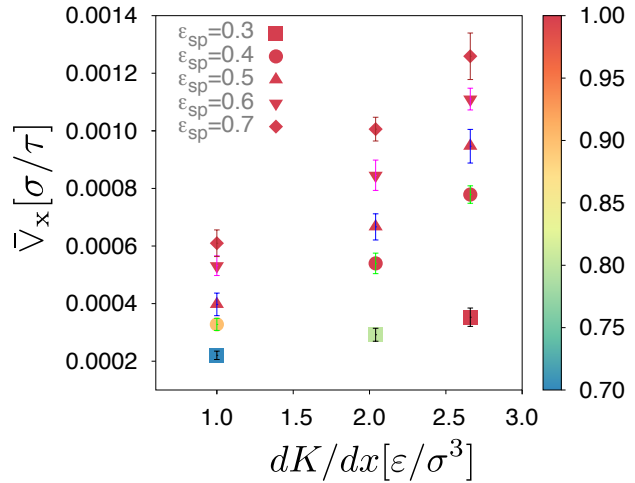}
\caption{\label{fig:3} The mean velocity $\bar{\vee}_{x}$ (similarly to
\ref{fig:2}) versus the gradient $dK/dx$ for various levels of attraction
between the droplet and the substrate, as indicated. The color bar indicates the
probability of durotaxis within the simulation time ($10^8$ MD steps).The
droplet is formed by $N_p=600$  chains.}
\end{figure}

Apparently, the results, displayed on Fig.~\ref{fig:2} indicate, that the mean
velocity of drops, $\bar{\vee}_{x}$, on three substrates with the {\em same} average
stiffness, increases steadily and proportionally to the magnitude of the
stiffness gradient $K'$. In addition, the higher attraction between
the droplet and the substrate leads to larger mean velocity of the droplet during
durotaxis irrespective of the magnitude of the stiffness gradient. Moreover,
the probability of durotaxis is lower than one  in the case of small gradients
and weak adhesion $\varepsilon_{sp}$. These findings represent the central
results of this investigation. A plot of our simulation data, cf.
Fig.~\ref{fig:3}, as a function of the gradient, $K'$, for different strength of
adhesion, $\varepsilon_{sp}$, and comparison with Fig.~\ref{fig:2} indicates
that the mean speed of durotaxis changes
linearly with the stiffness gradient, $K'$, and with the strength of attraction
between the droplet and the substrate, $\varepsilon_{\rm sp}$, namely, droplets
move faster on better wettable substrates. Therefore, our parameter exploration
for various cases, suggests that higher stiffness gradient and stronger affinity
between the droplet and the substrate result in more efficient durotaxis.

\begin{figure}[htb]
\includegraphics[scale=0.80]{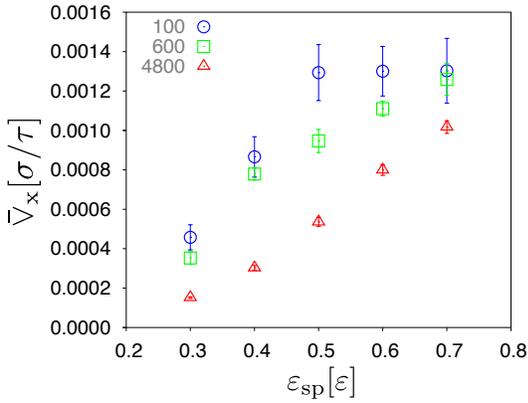}
\caption{\label{fig:4} The mean velocity $\bar{V}_{x}$ versus the parameter
$\varepsilon_{sp}$ that governs the adhesion to the substrate for droplets
containing $N_p=100,\; 600$, and $4800$ polymer chains with ten monomers. Here  
at the soft edge of the substrate $K_0=20$, and the stiffness gradient $K' =
2.7$. For all different cases of droplet size, the probability of
durotaxis is 100\%}.
\end{figure}

As is well known, the size of small (nano)droplets affects the resulting contact
angle and the degree of droplet adhesion on surfaces whenever the free energy of
the contact line becomes comparable to the surface free energy (surface
tension). One may, therefore, expect that durotaxis will also depend on droplet
size \cite{Chen}. We have investigated different drop sizes (see
Fig.~\ref{fig:4}) and we established that the bigger the droplet is, the less
efficient durotaxis becomes. Our observation holds for the whole range of
$\varepsilon_{\rm sp}$ values considered in this study. In particular, the
smallest droplet with ($N_{\rm p}=100$) exhibits the fastest durotaxis reaching
a threshold speed value when $\varepsilon_{\rm sp}>0.5$, whereas the largest
droplet ($N_p=4800$) exhibits a linear dependence with the parameter 
\begin{figure}[htb]
\includegraphics[scale=1.00]{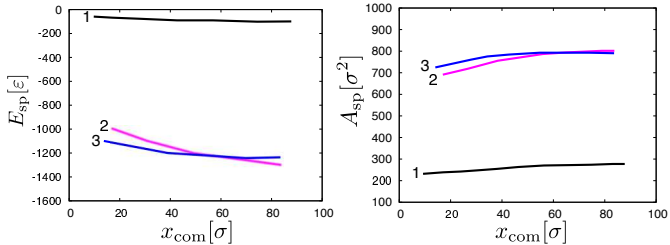}
\caption{\label{fig:5}  Interaction energy between droplet and substrate,
$E_{\rm sp}$ (left panel), and interfacial area, $A_{\rm sp}$ (right panel), as a function of the
centre-of-mass-position (COM) of the droplet, $x_{\rm com}$, for the cases $K_{\rm 0} = 20$, $K'
= 2.7,\; \varepsilon_{\rm sp}=0.30$ (1), $K_0=20,\; K'=2.7,\varepsilon_{\rm sp}= 0.70$ (2) and
$K_0=100,\; K'=1.0,\varepsilon_{\rm sp}= 0.70$ (3) for a drop with $N_p=600$. In the case of
optimal durotaxis, the droplet gains energy  $E_{\rm sp}$ faster as it moves to stiffer
parts of the substrate. In contrast, when this gain is negligible, no durotaxis
takes place. Here, substrate and droplet beads within the cutoff distance $2.5\sigma$ are considered as part of the interface.
The interfacial area has been calculated by projecting the droplet beads of the interface onto the z=0 layer and determining
the convex hull by using the qhull library. \cite{qhull} In this case the area of the convex hull is the interfacial area. }
\end{figure}
$\varepsilon_{\rm sp}$. We interpret this result as consequence of the increased
friction due to the larger area of contact with the substrate associated with
the big droplet. One should also note that this observation is at variance with
the results of Style {\it et al.} \cite{Style2013}, who found that a drop moves 
spontaneously toward softer parts of the substrate, once the contact angle
difference across the droplet, $\Delta \theta$, exceeds some critical value,
given that $\Delta \theta$ would be also larger for the bigger droplets. Our
results comply, however, with those of Chang {\it et al.} \cite{Chang2015} who
used similar architecture of the substrate as ours, which makes us believe that
the concrete implementation of varying degree of stiffness along the substrate
largely determines the outcome of durotaxis. 
\begin{figure}[htb]
\includegraphics[scale=0.25]{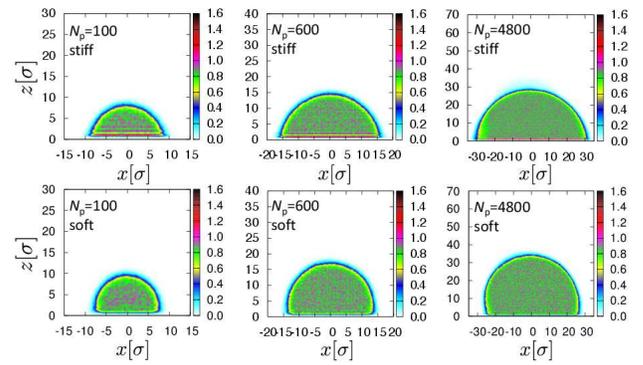}
 
\caption{\label{fig:6} The density profiles  on the $x-z$ plane for different
droplet sizes and substrate stiffness is illustrated, as indicated. For soft
substrate the stiffness is $K=276$, whereas for the soft substrates is $K=20$.
The figure illustrates the dependence of the contact angle with the droplet
size \cite{Theodorakis2015} and the substrate stiffness.}
\end{figure}

Based on our detailed analysis of a larger number of different properties and
systems, the driving force for durotaxis on a substrate with variable stiffness
stems from the possibility for the droplet to diminish its overall energy by
displacement to stiffer regions on the substrate, in agreement with previous
work in the context of a flake on a graphene layer \cite{Chang2015,Becton2016}.
Fig.~\ref{fig:5} illustrates  examples of systems with different durotaxis
efficiency. Comparing these results with those of Figures~\ref{fig:2} and \ref{fig:3}, we 
observe that larger variations in the interfacial energy $E_{\rm sp}$ with changing position $x$ lead to more
efficient durotaxis. 

\begin{figure}[htb]
\vspace{0.7cm}
\includegraphics[scale=0.30]{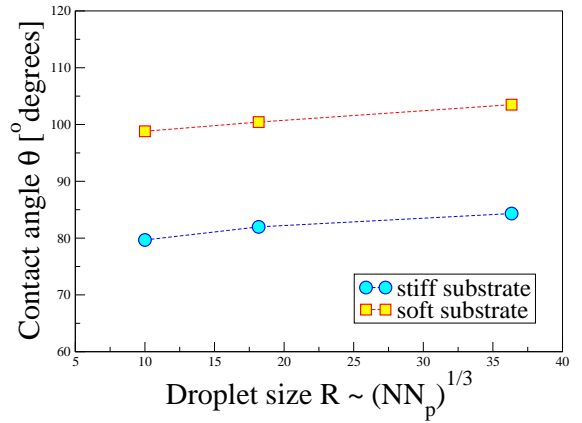}
\caption{Variation of the contact angle $\theta$ of sessile drops, cf.  
Fig.~\ref{fig:6}, with drop radius $R \propto (NN_p)^{1/3}$ on a soft, $K=20$, 
and stiff, $K=276$, substrates.  \label{fig:7} }
\end{figure}
 In this case, the 
droplet is able to establish a larger number of contacts with the substrate and 
get attracted stronger by the interface. This results in changes of many 
properties with the centre-of-mass displacement of the droplet, such as the 
radius of gyration of the droplet, the interfacial area between the droplet and 
the substrate, as well as the contact angle $\theta$. Analyzing the various 
contributions to the total energy of the system, it appears that minimization of 
the interfacial energy $E_{\rm sp}$ (expressed through $dE_{\rm sp}/dx_{com}$)
serves as the main driving force for efficient durotaxis.

Moreover, by exploring droplets of different size, and analyzing the respective
density profiles,  cf. Fig.~\ref{fig:6}), we find that larger droplets are
generally more hydrophobic than smaller ones for the same values of
$\varepsilon_{\rm sp}$ and degree of substrate stiffness. This indicates that as
the drop size decreases, a gain from negative line tension compensates
increasingly the energy expense related to the vapor-liquid surface of the drop
\cite{AMAM} as established earlier by Gretz \cite{Gretz}.

We also observe that the contact angle $\theta$ for droplets of different size 
is always smaller at stiffer substrates (whereby the stronger attraction to the 
substrate leads to layering effects) than it is for soft substrates (Figs.~\ref{fig:6} and \ref{fig:7}). As is 
evident from Fig.~\ref{fig:6}, the droplet does not ``immerse'' into the softer 
part of the substrate, but rather, irrespective of the stiffness, the substrate 
is able to support the droplet. One may, therefore, conclude that the 
interfacial energy, which is the governing force of durotaxis, depends on the 
size of the droplet, the substrate stiffness, as well as on the substrate 
wettability, whereby stronger stiffness gradients enhance the durotaxis.

\begin{figure}[htb]
\includegraphics[scale=0.90]{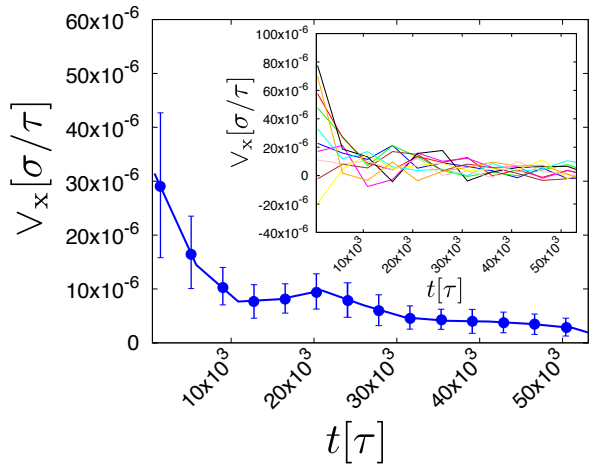}
\caption{\label{fig:8} The instantaneous velocity $\vee_{x}$ of droplets during
durotaxis versus elapsed time $t$ as an average over ten different trajectories
($V_x(t)$ for individual trajectories is shown in the inset). Here,
$N_p=600$, $K_0=20$, $K'=2.7$, and $\varepsilon_{sp}=0.7$. }
\end{figure}

\begin{figure}[htb]
\vspace{0.5cm}
\includegraphics[scale=0.25, angle=270]{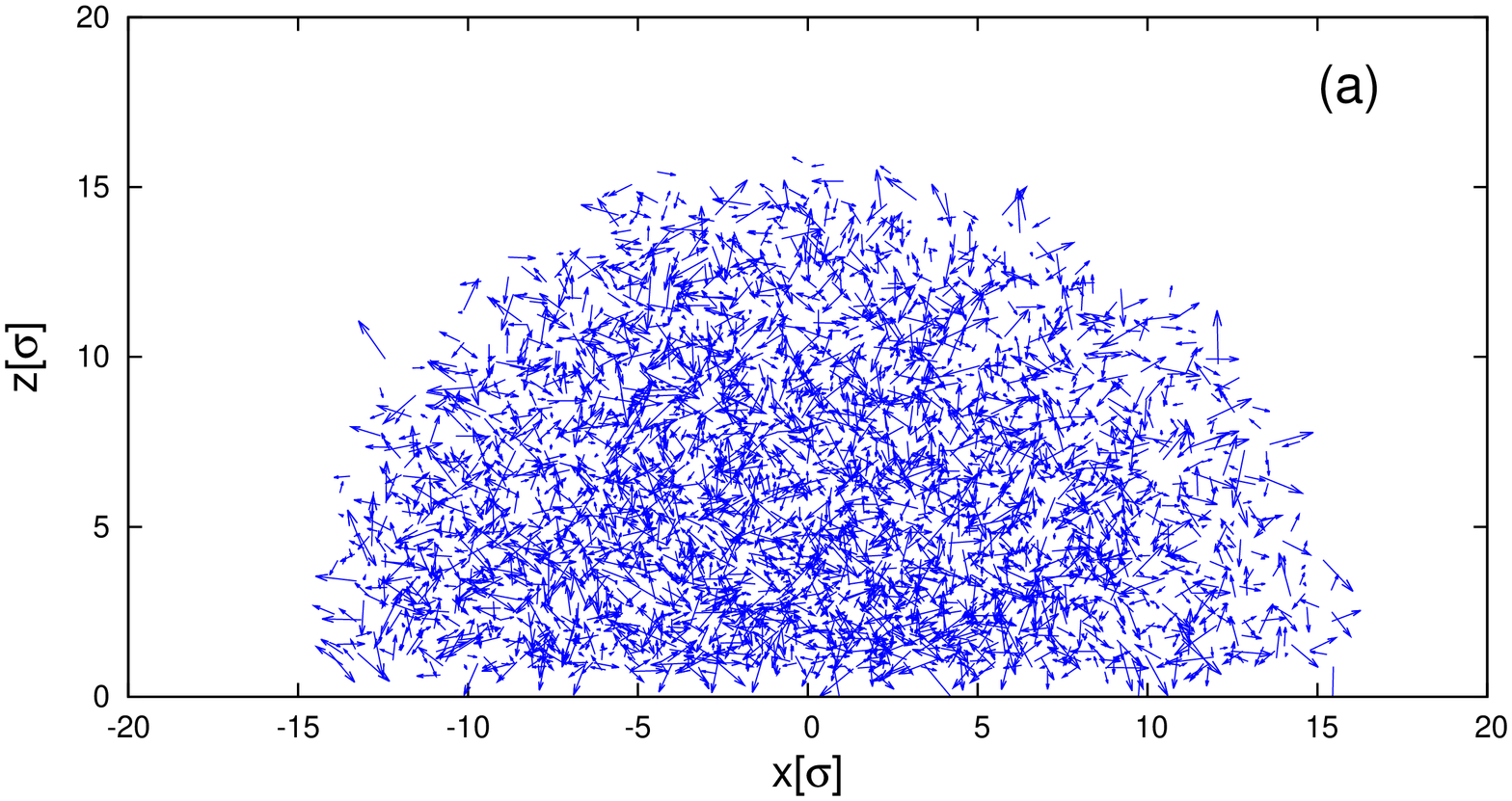}

\vspace{-1cm}
\includegraphics[scale=0.25, angle=270]{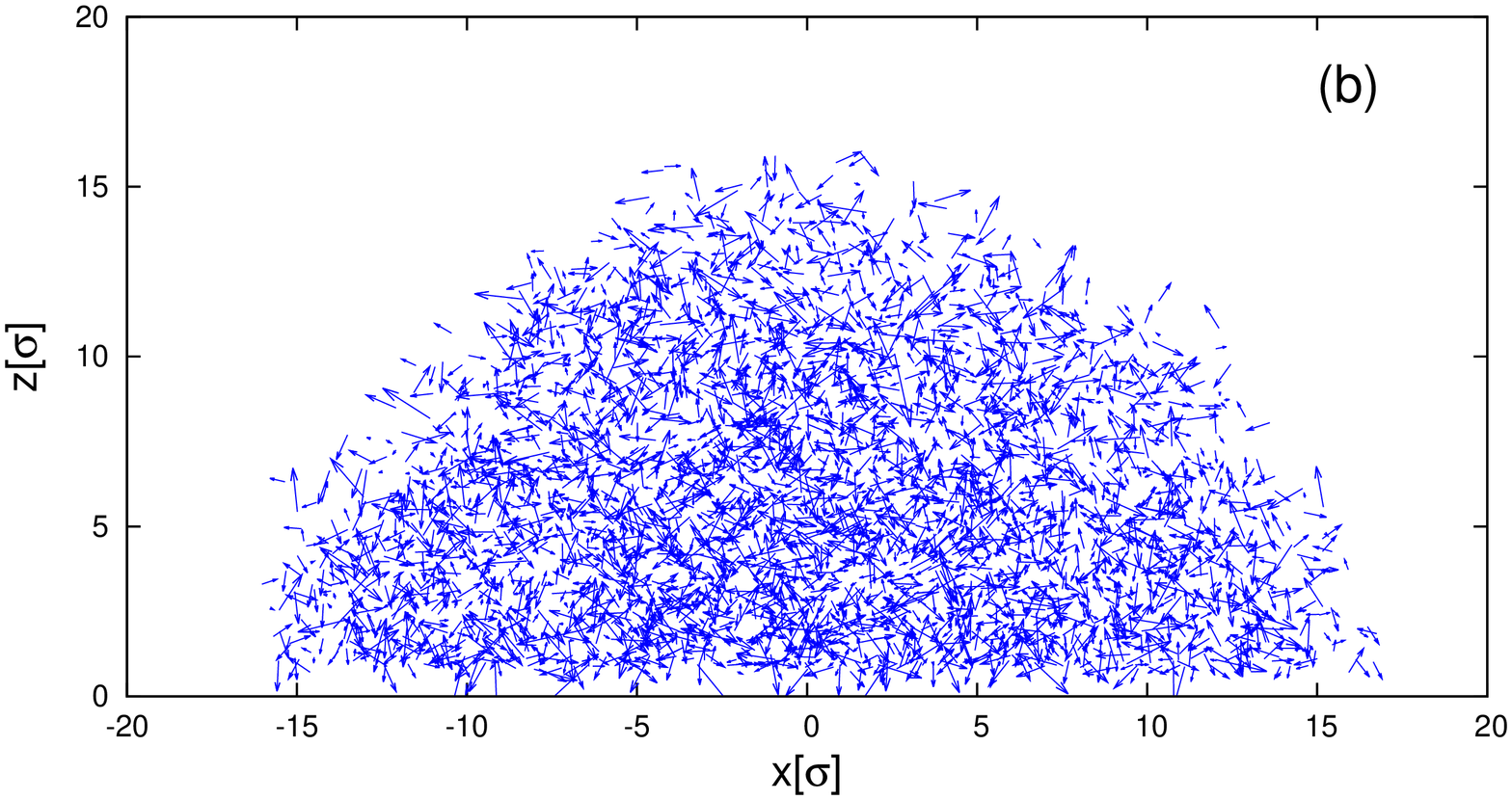}

\vspace{-1cm}
\includegraphics[scale=0.25, angle=270]{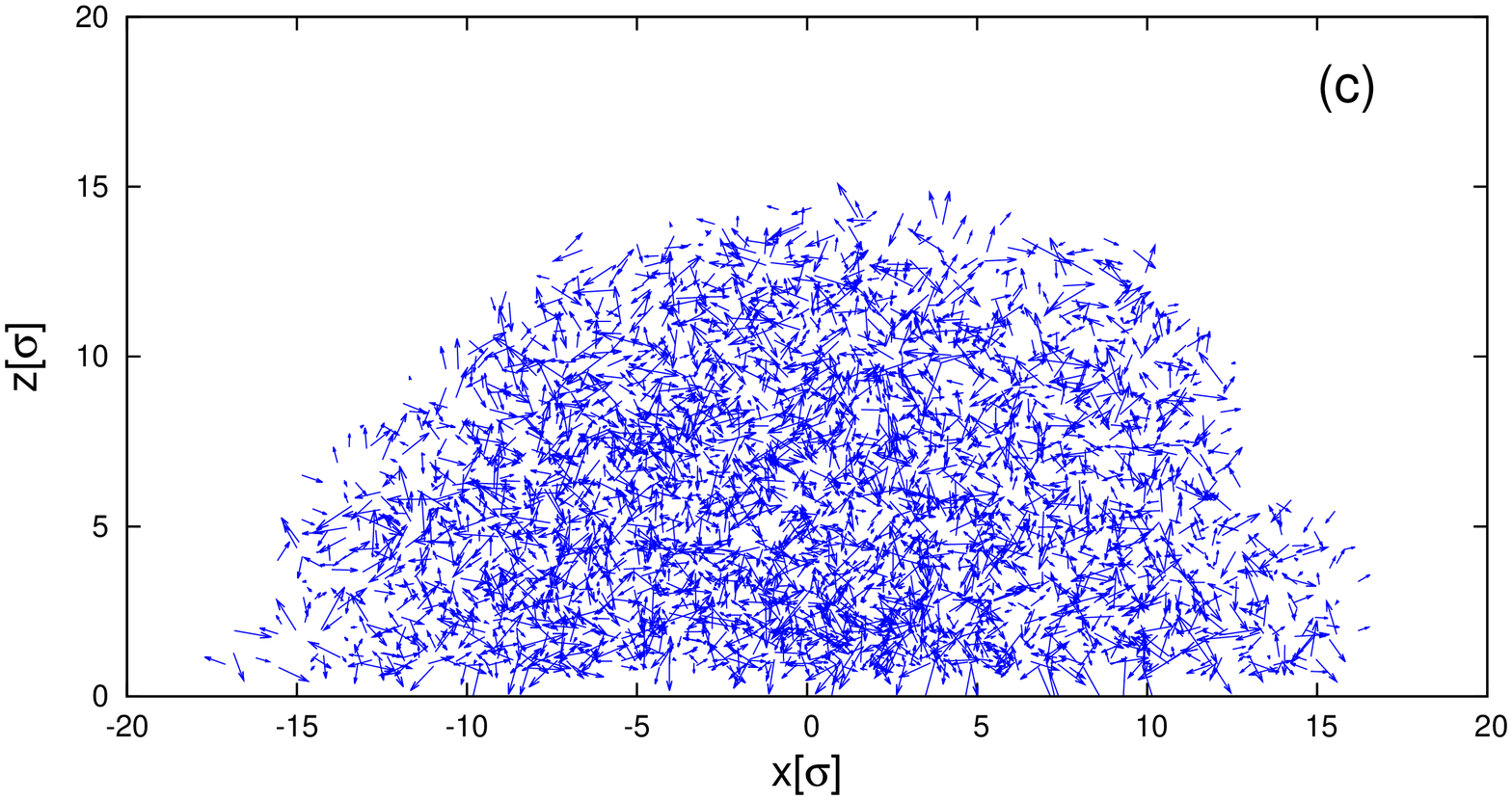}

\caption{\label{fig:9} Typical velocity vector fields in a cross section of the
droplet, projected to the  $x-z$ plane, at early (top, $t=2.5\times10^3\tau$),
(middle, $t=1.75\times10^4\tau$), and late time (bottom, $t=4.5\times10^4\tau$)
during durotaxis for the case of $N_p=600$, $K_0=20$, $\Delta K=10.66$, and
$\varepsilon_{sp}=0.7$.}
\end{figure}

Eventually, we also analyzed the instantaneous velocity of the droplets for
different cases. One of these cases is presented in Fig.~\ref{fig:8}. It shows
that the droplet does not maintain a constant velocity while crossing the
substrate along the gradient direction, but rather exhibits faster motion while
on the softer part of the substrate. Individual trajectories, however, are seen
to fluctuate strongly in their behavior in the course of the process. Overall,
the existence of stiffness gradient in the substrate competes with thermal
fluctuations, giving rise to a distorted translational movement of droplets to
stiffer parts of the substrate. This conclusion is further corroborated by
monitoring the velocity field of the droplet at various stages of durotaxis,
where no particular pattern for the motion of the polymer chains in the droplet
can be readily identified (Fig.~\ref{fig:9}). In fact, the droplet moves back
and forth during durotaxis with the stiffness gradient determining the direction
of the motion.

\section{\label{sec:summary} Concluding remarks}

In this study, we have investigated the durotaxis of a droplet onto a substrate, 
characterized by stiffness gradient in one direction. We observe spontaneous 
directional movement of the drops from softer to more rigid parts of the 
substrate, whereby on substrates with equal mean stiffness, durotaxis becomes 
progressively enhanced upon increasing the stiffness gradient strength. Thereby 
the average speed of directional motion increases up to a limiting maximal 
value. Furthermore, at given stiffness gradient, droplets with better adhesion 
to the substrate exhibit as a rule essentially faster movement toward the rigid 
part of the substrate. The latter finding is corroborated by the fact, that 
smaller droplets are also observed to better wet the 
substrate \cite{Theodorakis2015}, owing to a stronger interplay between contact 
line tension and surface tension of the vapor--polymer melt interface with 
decreasing size of the drop, exhibiting thus a more efficient durotaxis.

We interpret the observed motion of the droplet towards areas of larger 
stiffness of the substrate as manifestation of the tendency of the system to 
acquire  energetically favorable states, where the droplet establishes a larger 
number of contacts between substrate and polymer beads, gaining thus van der 
Waals contact energy of the droplet particles interaction with the particles of 
the substrate, in agreement with previous work which considered the durotaxis 
motion of a flake on a graphene layer \cite{Chang2015}. Our computer experiments 
demonstrate that this gain in contact energy increases with growing stiffness of 
the surface.

The present study explores the possibility of guided motion of droplets on 
variably stiff substrates with implications for nanofluidics, microfabrication, 
and coating. Our results establish the change in interfacial energy between  
droplet and substrate as the driving force for durotaxis, and we anticipate that 
this work will provide further insight into the mechanisms of nanoscale 
directional motion that has general implications in novel technologies and 
applications in biology and health care.

This research has been supported by the National Science Centre, Poland, under grant No.~2015/19/P/ST3/03541. This project has received funding from the European Union's Horizon 2020 research and innovation programme under the Marie Sk{\l}odowska--Curie grant agreement No. 665778. This research was supported in part by PLGrid Infrastructure.

\bibliography{durota}

\end{document}